\documentclass[11pt]{article}
\begin{document}

\begin{center}
\large \textbf{Thermodynamics and weak cosmic censorship conjecture in extended phase spaces of anti-de Sitter black holes with particles' absorption}
\end{center}

\begin{center}
Deyou Chen
\end{center}

\begin{center}
School of Science, Xihua University, Chengdu 610039, China

E-mail: deyouchen@hotmail.com
\end{center}

{\bf Abstract:} The thermodynamics and weak cosmic censorship conjecture in extended phase spaces of charged anti-de Sitter black holes describing the massive gravity are investigated by the absorptions of the scalar particle and fermion. The cosmological constant is seen as a pressure with a conjugate volume. The first law of thermodynamics is recovered. The second law of thermodynamics is violated in the extended phase space of the extremal black hole. For the near-extremal and extremal black holes, the validity of weak cosmic censorship conjecture is tested by evaluating the minimum values of the metric function $f$. It is found that they remain their near-extremity and extremity when the particles are absorbed.

\section{Introduction}

Spacetime singularities are formed at the end of gravitational collapse. At the singularities, the curvatures of spacetime tend to diverge and all laws of physics break down. To avoid destructions caused by singularities, Penrose put forward the weak cosmic censorship conjecture (WCCC) where naked singularities cannot be formed in a real physical process from regular initial conditions \cite{RP}. This conjecture shows that singularities are hidden behind event horizons of black holes and can not be observed by distant observers. Although its rationality is widely accepted, there is still a lack of complete proof, and people's interest is focused on the test of its validity.

A famous test is the Gedanken experiment proposed by Wald \cite{RMW}. In this experiment, a test particle with energy, large enough angular momentum and charge is thrown into a charged and rotating black hole to see whether the event horizon is destroyed. If the horizon is destroyed, there is no horizon solution in the metric function and the WCCC is invalid. For the Kerr-Newman black hole, the relation $M^2< a^2+Q^2$ shows the nonexistence of the horizon. His research showed that the extremal Kerr-Newman black hole could not be overcharged and overspun. Following this work, people studied the validity of the WCCC in the various spacetimes. Hubeny found that a near-extremal Reissner-Nordstrom black hole could be overcharged by absorbing a charged particle \cite{VEH}. The research of Jacobson and Sotiriou showed that a near-extremal Kerr black hole could be overspun \cite{JS}. Subsequent researches showed that the backreaction and self-force effects might prevent charged/rotating black holes from being overcharged/overspun \cite{SH,CB,CBSM,BCK1,BCK2,ZVPH}. Replacing the test particle with a test field, Semiz reviewed the validity of the WCCC in the dyonic Kerr-Newman black holes \cite{IS,GZT}. Other work about the validity of the WCCC are referred to \cite{CO,MS,RS1,SS,IST,SH2,GZ,DS1,HMS,KD,HS,RV,SW1,FGS,RMW2,ASZZ, GMZZ,BG1,CHS,YW,DYC1,DYC2,ZZ,ZHC,KJ}. Although a lot of work has been done, no consistent conclusion has been reached. Recently, Gwak discussed the validity of the WCCC in the extended phase space of the charged anti-de Sitter black hole \cite{BG2,BG3,GG}. He found that the entropy for the extremal and near-extremal black holes decreased under the charged particle absorption and the WCCC was still valid. A key point is finding the relation between the momentum and charge of the ingoing particle. This relation is derived by the Hamilton-Jacobi equation.

In this paper, we investigate the thermodynamics and WCCC in extended phase spaces of charged anti-de Sitter black holes by the absorptions of the scalar particle and fermion. This black hole solution is gotten from the massive gravity which is a infrared modification of Einstein's general relativity and propagates a massive spin-2 particle \cite{CGT,KH1,DV,CHPZ}. Their thermodynamical properties and phase structure have been discussed in \cite{CHPZ}. The cosmological constant used to be regarded as a fixed constant. Recently, it has been seen as a variable related to pressure $P=-\frac{\Lambda}{8\pi}$. One of the reasons is that the cosmological constant as a variable can reconcile the inconsistencies between the first law of black hole thermodynamics and the Smarr relation derived by the scaling method. The second reason is that the physical constants, such as Newtonian¡¯s constant, gauge coupling constants and the cosmological constant arise as vacuum expectation values and vary in more fundamental theories \cite{KRT1,KRT2,KM1,KM2}. In the extended phase spaces, people discovered a lot of new thermodynamic behaviors, such as the van der Waals liquid gas phase transition, reentrant phase transitions, tricritical points and so on. Therefore, our interest is focused on the extended phase spaces and the cosmological constant is induced a pressure with a conjugate volume. To discuss the thermodynamics, we assume that the final state is still a black hole. The first law of thermodynamics is recovered by the absorptions of the scalar particle and fermion, respectively. For the extremal black hole, the entropy decreases with the absorption of the particles, which violates the second law of thermodynamics. This result is different from that gotten in the normal phase spaces. To test the validity of the WCCC in the near-extremal and extremal black holes, we need to find out whether the horizon solutions exist, which is achieved by evaluating the minimum values of the metric function $f$. It is found that these black holes still remain their near-extremity and extremity when the particles are absorbed. Therefore, these black holes can not be overcharged.

The rest of this paper is organized as follows. In the next section, the solution of the charged anti-de Sitter black holes in the massive gravity is reviewed. The absorptions of the scalar particle and fermion are discussed in section 3 and 4, respectively. In section 5, the final state is assumed to be a black hole and the first law of thermodynamics is recovered by the absorptions of the scalar particle and fermion. In section 6, the validity of the WCCC in the near-extremal and extremal black holes is investigated by the minimum value of the function $f$. Section 7 is devoted to our discussion and conclusion.

\section{Charged anti-de Sitter black holes in the massive gravity}

The action for an $(n + 2)$-dimensional massive gravity is \cite{DV}

\begin{eqnarray}
\mathcal{S} &=& \frac{1}{16\pi}\int{dx^{n+2}\sqrt{-g}\left[R+\frac{n(n+1)}{l^2}-\frac{F^2}{4}+m^2\sum_{i=1}^4{c_iu_i(g,f)}\right]},
\label{eq2.1}
\end{eqnarray}

\noindent where $c_i$ are constants, $f$ is a fixed symmetric tensor called the reference metric, and $u_i$ are symmetric polynomials of the eigenvalues of the $(n+2)\times (n+2)$ matrix $\mathcal{K}_{\nu}^{\mu}=\sqrt{f^{\mu \alpha}g_{\alpha \nu}}$:

\begin{eqnarray}
u_1 &=& [\mathcal{K}], \quad u_2=[\mathcal{K}]^2-[\mathcal{K}^2], \quad u_3=[\mathcal{K}]^3-3[\mathcal{K}][\mathcal{K}^2]+2[\mathcal{K}^3],\nonumber\\
u_4 &=& [\mathcal{K}]^4-6[\mathcal{K}^2][\mathcal{K}]^2+8[\mathcal{K}^3][\mathcal{K}]+3[\mathcal{K}^2]^2-6[\mathcal{K}^4].
\label{eq2.2}
\end{eqnarray}

\noindent The square root in $\mathcal{K}$ denotes $(\sqrt{A})_{\nu}^{\mu}(\sqrt{A})_{\lambda}^{\nu}=A_{\lambda}^{\mu}$ and $[\mathcal{K}]= \mathcal{K}_{\mu}^{\mu}$.

The solution of the static black hole with the spacetime metric and reference metric is given by \cite{CHPZ}

\begin{eqnarray}
ds^2 = -fdt^2 + \frac{1}{f}dr^2 + r^2 h_{ij}dx^idx^j,
\label{eq2.3}
\end{eqnarray}

\begin{eqnarray}
f_{\mu\nu}=diag(0,0,c_0h_{ij}),
\end{eqnarray}

\noindent where $c_0$ is a positive constant, and $h_{ij}dx^idx^j$ is the line element for an Einstein space with the constant curvature $n(n-1)k$. $k = 1, 0,$ or $-1$ denote a spherical, Ricci flat or hyperbolic topology black hole horizon, respectively. The metric function $f$ is

\begin{eqnarray}
f&=& k+\frac{r^2}{l^2}-\frac{16\pi M}{n\Omega_nr^{n-1}}+\frac{(16\pi Q)^2}{2n(n-1)\Omega_n^2}+\frac{c_0c_1m^2r}{n}+c_0^2c_2m^2\nonumber\\
&&+\frac{(n-1)c_0^3c_3m^2}{r} +\frac{(n-1)(n-2)c_0^4c_4m^2}{r^2},
\label{eq2.4}
\end{eqnarray}

\noindent where $M$ is the black hole mass, $Q$ is related to the charge and $\Omega_n$ is the space volume spanned by coordinates $x_i$. The cosmological constant $\Lambda$ is related to the constant $l^2$ and induced to the pressure $P=-\frac{\Lambda}{8\pi}=\frac{n(n+1)}{16\pi l^2}$. The electromagnetic potential is $A_{\mu}=-\frac{16\pi Q}{(n-1)\Omega_nr^{n-1}}dt$. Our interest is been focused on a four-dimensional spherically symmetric spacetime. One reason is the $AdS_4/CFT_3$ correspondence. It describes the correspondence between the quantum gravity theory in the four-dimensional anti-de Sitter spacetime and the conformal field theory on the boundary of this spacetime. Various tests and researches on the $AdS_4/CFT_3$ correspondence can be found in \cite{HS1,CGKP,NB,HKPT}. The four-dimensional case implies $k=1$, $n=2$, $A_{\mu}= -\frac{4Q}{r}dt$, $ P=\frac{3}{8\pi l^2}$ and $ r^2 h_{ij}dx^idx^j = r^2 (d\theta^2+ \sin^2\theta d\varphi^2)$. We set $c_3=c_4=0$ and get

\begin{eqnarray}
f&=& 1+\frac{r^2}{l^2}-\frac{2 M}{r}+\frac{(2 Q)^2}{r^2}+\frac{c_0c_1m^2r}{2}+c_0^2c_2m^2.
\label{eq2.5}
\end{eqnarray}

\noindent The event horizon $r_+$ is derived from $f=0$. The black hole mass is expressed as $M =\frac{r_+}{2}+\frac{r_+^3}{2l^2}+ \frac{2Q^2}{r_+}+ \frac{c_0c_1m^2r_+^2}{4}+ \frac{c_0^2c_2m^2r_+}{2}$. The Kretschmann scalar is

\begin{eqnarray}
K &=& \frac{896Q^4}{r^8}-\frac{384MQ^2}{r^7}+\frac{16(3M^2+2c_0^2c_2m^2Q^2)}{r^6}-\frac{16(c_0^2c_2m^2M+c_0c_1m^2Q^2)}{r^5}\nonumber\\
&&+\frac{4c_0^4c_2^2m^4}{r^4}+\frac{4c_0^3c_1c_2m^4}{r^3} +\frac{2(4c_0^2c_2m^2+c_0^2c_1^2l^2m^4)}{l^2r^2}+ \frac{12c_0c_1m^2}{l^2r}+\frac{24}{l^4}.
\end{eqnarray}

\noindent The curvature singularity is located at the solution of $K\to \infty$. When $Q=m=0$, the above scalar is reduced to $\frac{24}{l^4}+\frac{48M^2}{r^6}$, which is the Kretschmann scalar of the Schwarzschild anti-de Sitter black hole.

The entropy, volume and Hawking temperature are

\begin{eqnarray}
S =\pi r_+^2, \quad  V = \frac{4\pi r_+^3}{3}, \quad
T= \frac{1}{2\pi}\left(\frac{r_+}{l^2}-\frac{M}{r_+^2}-\frac{4 Q^2}{r_+^3}+\frac{c_0c_1m^2}{4}\right),
\label{eq2.6}
\end{eqnarray}

\noindent respectively. The electromagnetic potential measured at infinity with respect to the horizon is $\Phi =\frac{4Q}{r_+}$. These thermodynamic quantities obey the first law of thermodynamics in the extended phase space

\begin{eqnarray}
dM =TdS +\Phi dQ+ V dP.
\label{eq2.7}
\end{eqnarray}

\noindent When the cosmological constant is regarded as a constant, the above equation is reduced to the first law $dM =TdS +\Phi dQ$. In this paper, we regard the cosmological constant as a variable related to the pressure. Therefore, the first law (\ref{eq2.7}) is our concern. It is recovered by the absorptions of the scalar particles and fermions.

\section{Scalar particle absorption}

In this section, we discuss a scalar particle absorption by the four-dimensional spherically symmetric black hole in the massive gravity. In the curved spacetime, the motion of the charged scalar particle obeys the Klein-Gordon equation

\begin{eqnarray}
\frac{1}{\sqrt{-g}}\left(\frac{\partial}{\partial x^{\mu}}-\frac{iq}{\hbar}A_{\mu}\right)\left[\sqrt{-g}g^{\mu\nu}\left(\frac{\partial}{\partial x^{\nu}}-\frac{iq}{\hbar}A_{\nu}\right)\right]\Psi_S-\frac{m^2}{\hbar^2}\Psi_S =0,
\label{eq3.1}
\end{eqnarray}

\noindent where $m$ and $q$ are the mass and charge of the particle, respectively. Using the WKB approximation, we write the wave function as the following form

\begin{eqnarray}
\Psi_S= exp\left(\frac{i}{\hbar}I+I_1+\mathcal{O}(\hbar)\right).
\label{eq3.2}
\end{eqnarray}

\noindent The above ansatz was adopted in the discussions of the tunneling radiation of scalar particles \cite{UGRS,EGLSY,SG}. Inserting the above wave function and the contravariant metric components into the Klein-Gordon equation and taking the leading contribution of $\hbar$ yield

\begin{eqnarray}
f^{-1}(\partial_{t}I-qA_t)^2-f(\partial_{r}I)^2-\frac{1}{r^2}(\partial_{\theta}I)^2-\frac{1}{r^2\sin^2\theta}(\partial_{\varphi}I)^2+m^2=0.
\label{eq3.3}
\end{eqnarray}

\noindent Taking into account the symmetries of the spacetime, we carry out the separation of variables in the action

\begin{eqnarray}
I=-\omega t+W(r)+S(\theta,\varphi).
\label{eq3.4}
\end{eqnarray}

\noindent Plugging the above separated action into Eq.(\ref{eq3.3}), we get

\begin{eqnarray}
\partial_{r}W=\pm \frac{\sqrt{\left(\omega -\frac{4qQ}{r}\right)^2+\left[m^{2}-\frac{1}{r^2}(\partial_{\theta}S)^2-\frac{1}{r^2\sin^2\theta}(\partial_{\varphi}S)^2\right]f}}{f},
 \label{eq3.5}
\end{eqnarray}

\noindent In the work of Gwak \cite{BG3}, the action was gotten by the Hamilton-Jacobi equation and the positive sign was selected. Here, we still select the positive sign in the above equation. It is known that $p_{r}=\partial_{r}I=\partial_{r}W$. Therefore, we have

\begin{eqnarray}
p^r=g^{rr}p_r=\sqrt{\left(\omega -\frac{4qQ}{r}\right)^2+\left[m^{2}-\frac{1}{r^2}(\partial_{\theta}S)^2-\frac{1}{r^2\sin^2\theta}(\partial_{\varphi}S)^2\right]f}.
\label{eq3.6}
\end{eqnarray}

\noindent Near the event horizon where $f\to 0$, the above equation is reduced to

\begin{eqnarray}
p^r=\omega -q\Phi,
\label{eq3.7}
\end{eqnarray}

\noindent where $\Phi=\frac{4Q}{r_+}$ is the electromagnetic potential and $p^r$ is a positive value.  This is a relation between the momentum, energy and charge of the ingoing particle. For a special case $\omega =q\Phi$, the energy of the black hole does not change. When $\omega <q\Phi$, the energy of the black hole flows out the horizon and the supperradiation occurs. In this paper, we assume that the supperradiation does not occur. This implies $\omega \geq q\Phi$. The relation (\ref{eq3.7}) is adopted to discuss the black hole's thermodynamics in section 5 and recovered by the fermion absorption in the next section.

\section{Fermion absorption}

In curved spacetime, the motion of a charged fermion obeys the Dirac equation

\begin{eqnarray}
i\gamma^{\mu}\left(\partial_{\mu}+\Omega_{\mu}-\frac{i}{\hbar}qA_{\mu}\right)\Psi_{F}+\frac{m_0}{\hbar}\Psi_{F}=0,
\label{eq4.1}
\end{eqnarray}

\noindent where $m_0$ and $q$ are the mass and charge of the fermion, respectively. $\Omega _\mu \equiv\frac{i}{2}\omega _\mu\, ^{a b} \Sigma_{ab}$, $\omega _\mu\, ^{ab}$ is the spin connection defined by the ordinary connection and the tetrad $e^\lambda\,_b$

\begin{eqnarray}
\omega_\mu\,^a\,_b=e_\nu\,^a e^\lambda\,_b \Gamma^\nu_{\mu\lambda}
-e^\lambda\,_b\partial_\mu e_\lambda\,^a.
\label{eq4.2}
\end{eqnarray}

\noindent The Greek indices are raised and lowered by the curved metric $g_{\mu\nu}$. The Latin indices are governed by the flat metric $\eta_{ab}$. To construct the tetrad, we use the following definitions,

\begin{equation}
g_{\mu\nu}= e_\mu\,^a e_\nu\,^b \eta_{ab},\hspace{5mm} \eta_{ab}=
g_{\mu\nu} e^\mu\,_a e^\nu\,_b, \hspace{5mm} e^\mu\,_a e_\nu\,^a=
\delta^\mu_\nu, \hspace{5mm} e^\mu\,_a e_\mu\,^b = \delta_a^b.
\label{eq4.3}
\end{equation}

\noindent In the definition of $\Omega_\mu$, $\Sigma_{ab}$'s are the Lorentz spinor generators defined by

\begin{equation}
\Sigma_{ab}= \frac{i}{4}\left[ {\gamma ^a ,\gamma^b} \right],
\hspace{5mm} \{\gamma ^a ,\gamma^b\}= 2\eta^{ab}. \label{eq4.4}
\end{equation}

\noindent Therefore, it is readily to construct the $\gamma^\mu$'s in curved spacetime as

\begin{eqnarray}
\gamma^\mu = e^\mu\,_a \gamma^a, \hspace{7mm} \left\{ {\gamma ^\mu,\gamma ^\nu } \right\} = 2g^{\mu \nu }. \label{eq4.5}
\end{eqnarray}

For a fermion with spin $1/2$, its wave function must be well described with both spin up and spin down. We first describe the wave function with spin up. It takes the form as

\begin{eqnarray}
\Psi_{F}=\left(\begin{array}{c}
A\\
0\\
B\\
0
\end{array}\right)\exp\left(\frac{i}{\hbar}I\left(t,r,\theta,\varphi\right)\right),
\label{eq4.6}
\end{eqnarray}

\noindent where $A$, $B$ and the action $I$ are functions of $t, r , \theta , \phi$. To solve the wave function, we should construct the $\gamma^\mu$ matrices. There are many choices to construct them. For the metric (\ref{eq2.3}), we first chose the tetrad

\begin{eqnarray}
e_\mu\,^a = \rm{diag}\left(\sqrt f, 1/\sqrt f, r,r\sin\theta\right).
\end{eqnarray}

\noindent Then the $\gamma^\mu$ matrices are

\begin{eqnarray}
\gamma^{t}=\frac{1}{\sqrt{f\left(r\right)}}\left(\begin{array}{cc}
i & 0\\
0 & -i
\end{array}\right), &  & \gamma^{\theta}=r\left(\begin{array}{cc}
0 & \sigma^{1}\\
\sigma^{1} & 0
\end{array}\right),\nonumber \\
\gamma^{r}=\sqrt{f\left(r\right)}\left(\begin{array}{cc}
0 & \sigma^{3}\\
\sigma^{3} & 0
\end{array}\right), &  & \gamma^{\varphi}=r\sin\theta\left(\begin{array}{cc}
0 & \sigma^{2}\\
\sigma^{2} & 0
\end{array}\right).
\label{eq4.7}
\end{eqnarray}

\noindent In the above equations, $\sigma ^i$'s are the Pauli matrices given by

\begin{eqnarray}
\sigma^{1}=\left(\begin{array}{cc}
0 & 1\\
1 & 0
\end{array}\right), \quad \sigma ^{2}=\left(\begin{array}{cc}
0 & -i\\
i & 0
\end{array}\right), \quad \sigma ^{3}=\left(\begin{array}{cc}
1 & 0\\
0 & -1
\end{array}\right).
\label{eq4.71}
\end{eqnarray}

\noindent Inserting the gamma matrices and wave function into the Dirac equation (\ref{eq4.1}), applying the WKB approximation, and keeping only the leading order of $\hbar$, we get the following equations

\begin{eqnarray}
-iA\frac{1}{\sqrt{f}}(\partial_{t}I - qA_t)-B\sqrt{g}\partial_{r}I+Am_0 = 0,
\label{eq4.8}
\end{eqnarray}

\begin{eqnarray}
-iB\frac{1}{\sqrt{f}}(\partial_{t}I - qA_t)-A\sqrt{g}\partial_{r}I+Bm_0 = 0,
\label{eq4.9}
\end{eqnarray}

\begin{eqnarray}
 A\left[r\partial _{\theta} I+ir\sin\theta\partial _ {\varphi}I\right]= 0,
\label{eq4.10}
\end{eqnarray}

\begin{eqnarray}
B\left[r\partial _{\theta} I+ir\sin\theta\partial _ {\varphi}I\right]= 0,
\label{eq4.11}
\end{eqnarray}

\noindent There are four equations, our attention is focused on the first two equations, since the radial action is determined by them. The angular action does not affect our result, therefore, we don't need to know its expression. Considering the question we addressing and following the standard process, we carry out separation of variables

\begin{eqnarray}
I = -\omega t + W(r) + \Theta (\theta ,\varphi ), \label{eq4.12}
\end{eqnarray}

\noindent where $\omega$ is the energy of the ingoing fermion. Substituting equation (\ref{eq4.12}) into equations (\ref{eq4.8}), (\ref{eq4.9}) and canceling $A$ and $B$, we get

\begin{eqnarray}
f^2\left(\partial_{r}W\right)^{2}-\left(\omega -\frac{4qQ}{r}\right)^2-m_0^{2}f=0,
\label{eq4.13}
\end{eqnarray}

\noindent which yields

\begin{eqnarray}
\partial_{r}W=\pm \frac{\sqrt{\left(\omega -\frac{4qQ}{r}\right)^2+m_0^{2}f}}{f}.
\label{eq4.14}
\end{eqnarray}

\noindent Using $p_{r}=\partial_{r}I=\partial_{r}W$ and following the derivation in the above section and the work of Gwak \cite{BG3}, we still select the positive sign in the above equation and get

\begin{eqnarray}
p^r=g^{rr}p_r=\sqrt{\left(\omega -\frac{4qQ}{r}\right)^2+m_0^{2}f}.
\label{eq4.15}
\end{eqnarray}

\noindent We focus our attention on the ingoing wave function near the event horizon where $f\to 0$. The above equation is simplified to

\begin{eqnarray}
p^r=\omega -q\Phi,
\label{eq4.16}
\end{eqnarray}

\noindent which is full in consistence with that derived in Eq.(\ref{eq3.7}). Therefore, the relation (\ref{eq3.7}) is recovered by fermion absorption. The above discussion is about the spin up state. For the spin down state, the calculation and discussion are parallel. The same result can be gotten. We do not discuss it in detail here.

\section{Thermodynamics and particles' absorption}

In this section, we use the relations (\ref{eq3.7}) or (\ref{eq4.16}) to discuss the thermodynamics of the black hole. Therefore, the ingoing particle can be either the scalar particle or the fermion. When the black hole absorbs the particle with energy $\omega$ and charge $q$, its internal energy and charge with the pressure and radius are changed. The initial state of the black hole is represented by $(M,Q,P,r_+)$, and the final state is represented by $(M+dM,Q+dQ,P+dP,r_++dr_+)$. $dM$, $dQ$, $dP$ and $dr_+$ denote the increases of the mass, charge, pressure and radius, respectively. The functions $f(M+dM,Q+dQ,P+dP,r_++dr_+)$ and $f(M,Q,P,r_+)$ satisfy

\begin{eqnarray}
f(M+dM,Q+dQ,P+dP,r_++dr_+)&=&f(M,Q,P, r_+)+\left.\frac{\partial f}{\partial M}\right|_{r=r_+}dM + \left.\frac{\partial f}{\partial Q}\right|_{r=r_+}dQ  \nonumber\\
&&+ \left.\frac{\partial f}{\partial P}\right|_{r=r_+}dP+ \left.\frac{\partial f}{\partial r}\right|_{r=r_+}dr_+,
\label{eq5.1}
\end{eqnarray}

\noindent where

\begin{eqnarray}
\left.\frac{\partial f}{\partial M}\right|_{r=r_+}&=&-\frac{2}{r_+},\quad \left.\frac{\partial f}{\partial Q}\right|_{r=r_+}=\frac{8Q}{r_+^2}, \nonumber\\ \left.\frac{\partial f}{\partial P}\right|_{r=r_+}&=&\frac{8\pi r_+^2}{3}, \quad
\left.\frac{\partial f}{\partial r}\right|_{r=r_+} =4\pi T.
\label{eq5.2}
\end{eqnarray}

\noindent Both of the initial and final states are assumed to be black holes, which leads to $f(M+dM,Q+dQ,P+dP,r_++dr_+)=f(M,Q,P,r_+)= 0$. In the extended phase space, the black hole mass is regarded as an enthalpy which relates to the internal energy as $U=M-PV$. The increases of the energy and charge of the black hole are related to the energy and charge of the ingoing particle, namely, $dU=\omega$ and $dQ=q$. Thus, Eq.(\ref{eq4.16}) becomes

\begin{eqnarray}
p^r +\Phi dQ=dU=d(M-PV).
 \label{eq5.3}
\end{eqnarray}

\noindent This relation is different from that in the normal phase space where the increase of energy is related to the black hole's mass. Combining the above equation with Eq.(\ref{eq5.1}) and Eq.(\ref{eq5.2}), we get

\begin{eqnarray}
dr_+=\frac{p^r}{(2\pi T-4\pi r_+ P)r_+}.
\label{eq5.4}
\end{eqnarray}

\noindent Using the relation between the entropy and radius, $S=\pi r_+^2$, yields the variation of the entropy

\begin{eqnarray}
dS=2\pi r_+dr_+=\frac{p^r}{ T-2 r_+ P}.
\label{eq5.5}
\end{eqnarray}

\noindent The variations of the radius and entropy are determined by the value of $T-2 r_+ P$. When $T>2 r_+ P$, the radius and entropy increase. $T<2 r_+ P$ implies the decreases of the radius and entropy. For the extremal black hole where $T=0$, one can get $dS<0$. The second law of thermodynamics implies that the black hole entropy increases in an irreversible process. The absorption of the particle by the black hole is an irreversible process, therefore, $dS<0$ violates the second law. This reusult is different from that in the normal phase space.

We further discuss the black hole's thermodynamics. From Eq.(\ref{eq5.3}) and Eq.(\ref{eq5.5}), we get

\begin{eqnarray}
dM=TdS+VdP+\Phi dQ.
\label{eq5.6}
\end{eqnarray}

\noindent The first law of thermodynamics is recovered by the absorptions of the scalar particle and fermion. This result is inevitable because both the initial and final states are assumed to be black holes. Therefore, the black hole thermodynamics were discussed through the absorptions of the scalar particle and fermion at the same time. In the next section, we discuss the validity of the WCCC in the near-extremal and extremal black holes.

\section{WCCC in the near-extremal and extremal black holes}

For the near-extremal and extremal black holes, it is difficult to judge the violation of the WCCC since we do not know whether their final states are black holes after absorptions of test particles with energy and large enough charge. Therefore, we should find out their states. If there is a positive real root in the function $f$, their final states are black holes and the WCCC is valid. Otherwise, the WCCC is violated.

The existence of the root shows that there is a minimum value for the function $f$. At this value, there are

\begin{eqnarray}
f(M,Q,P,r_0)&=&1-\frac{2M}{r_0}+\frac{Q^2}{r_0^2}+\frac{r_0^2}{l^2}\leq 0, \nonumber\\
\left.\frac{\partial f(M,Q,P,r)}{\partial r}\right|_{r=r_0}&=&0,
\label{eq6.1}
\end{eqnarray}

\noindent where $r_0$ is the location corresponding to the minimum value. $f(M,Q,P,r_0)$ is equal to zero for the extremal black hole and less than zero for the near-extremal black hole. $(M,Q,P,r_0)$ and $(M+dM,Q+dQ,P+dP,r_0+dr_0)$ represent the initial state and the finial state, respectively. We express the function $f$ at the final state in term of that at the initial state

\begin{eqnarray}
f(M+dM,Q+dQ,P+dP,r_0+dr_0)&=&f(M,Q,P,r_0)+\left.\frac{\partial f}{\partial M}\right|_{r=r_0}dM \nonumber\\
&& + \left.\frac{\partial f}{\partial Q}\right|_{r=r_0}dQ + \left.\frac{\partial f}{\partial P}\right|_{r=r_0}dP,
\label{eq6.2}
\end{eqnarray}

\noindent where the second equation in Eq.(\ref{eq6.1}) was used  and

\begin{eqnarray}
\left.\frac{\partial f}{\partial M}\right|_{r=r_0}=-\frac{2}{r_0},\quad \left.\frac{\partial f}{\partial Q}\right|_{r=r_0}=\frac{8Q}{r_0^2},\quad \left.\frac{\partial f}{\partial P}\right|_{r=r_0}=\frac{8\pi r_0^2}{3}.
\label{eq6.3}
\end{eqnarray}

\noindent We first evaluate the minimum value in the extremal black hole where $r_+=r_0$. Using Eq.(\ref{eq5.3}), Eq.(\ref{eq6.2}) and Eq.(\ref{eq6.3}) yields

\begin{eqnarray}
f(M+dM,Q+dQ,P+dP,r_0+dr_0)=0.
\label{eq6.4}
\end{eqnarray}

\noindent It shows that when the particle is absorbed, the final state of the extremal black hole in the extended phase space is still an extremal with new mass and charge. Therefore, the existence of the event horizon ensures that the singularity is not naked in this black hole.

For the near-extremal black hole, we order $|f(M,Q,P,r_0)| \ll 1$ to ensure its near-extremity. It is difficult to evaluate the minimum value. We let $r_0=r_+-\epsilon$, where $0<\epsilon\ll 1$. Eq.(\ref{eq6.2}) is rewritten in terms of $\epsilon$ and $r_+$ as

\begin{eqnarray}
f(M+dM,Q+dQ,P+dP,r_0+dr_0)&=&f(M,Q,P,r_0)-\frac{2Tp^r}{(T-2 r_+ P)r_+}\left(1+\frac{2\epsilon}{r_+}\right)\nonumber\\
&&+\frac{2\epsilon}{r_+^2}dM+0(\epsilon)\nonumber\\
&=&f(M,Q,P,r_0)+\frac{2\epsilon}{r_+^2}dM+\mathcal{O}(\epsilon)<0.
\label{eq6.5}
\end{eqnarray}

\noindent In the above equation, $\mathcal{O}(\epsilon)$ is the higher order of $\epsilon$, which can be neglected. Since $r_+\sim l$, $l=\sqrt{-3\Lambda^{-1}}\gg1$, and $\epsilon\ll1$, $\frac{2\epsilon}{r_+^2}dM$ can also be neglected. The near-extremity leads to $T\to 0$. The second item on the right hand of the first equal sign of the above equation is neglected. The result shows that when the near-extremal black hole absorbs the particle, its final state is still a near-extremal black hole with new mass and charge. Therefore, the singularity is hidden behind the event horizon and the WCCC is valid in this black hole.

\section{Discussion and Conclusion}

In this paper, we investigated the thermodynamics and WCCC in the extended phase spaces of the anti-de Sitter black holes in the massive gravity by the absorptions of the scalar particle and fermion. The cosmological constant was regarded as a pressure with a conjugate volume. The final state was assumed to be a black hole. The first law of thermodynamics was recovered by the absorptions of the particles. The second law of thermodynamics shows that the black hole entropy increases in an irreversible process. We found that the second law was violated in the extended phase space of the extremal black hole. For the extremal and near-extremal black holes, we judged the existences of the event horizons by evaluating the minimum values of the function $f$. The minimum values of the function $f$ remain unchanged after the particles with energy and large enough charge are absorbed, which shows that both of the extremal and near-extremal black holes can not be overcharged. This result is in accordance with that gotten by Gwak \cite{BG3}. In Gwak's work, he derived the relation between the action and momentum by the Hamilton-Jacobi equation.

In fact, inserting the wave function (\ref{eq3.2}) into the Klein-Gordon equation (\ref{eq3.1}) and keeping only the leading order terms in $\hbar$ yield the Hamilton-Jacobi equation

\begin{eqnarray}
g^{\mu\nu}\left(\frac{\partial I}{\partial {x^\mu}}-qA_\mu\right)\left(\frac{\partial I}{\partial {x^\nu}}-qA_\nu\right)+m^2=0.
\label{eq7.1}
\end{eqnarray}

\noindent On the other hand, the Hamilton-Jacobi equation can be obtained from the Dirac equation \cite{EL}. The multiplication of (\ref{eq4.1}) yields the second-order partial derivative equation. We write the wave function as $\Psi_{F}=\Psi_{0}e^{\frac{i}{\hbar}I}$, where $\Psi_{0}$ is a position-dependent spinor. The Hamilton-Jacobi equation is derived by inserting the function into the second-order partial derivative equation and keeping only the leading order terms in $\hbar$. Therefore, our result is naturally consistent with that of Gwak.

\vspace*{2.0ex}
\textbf{Acknowledgments}
I am very grateful for Professor Y.P. Hu for his useful discussion.

\end{document}